\documentclass{appolb}
\usepackage{graphicx}
\usepackage{amsmath}

\begin{document}

\title{Gossip in random networks
\thanks{Presented at the 2nd Polish Seminar on Econo- and Sociophysics, Cracow, 21-22 April 2006}
}
\author{K. Malarz
\footnote{\tt http://home.agh.edu.pl/malarz/}
\address{Faculty of Physics and Applied Computer Science,
AGH University of Science and Technology,
al. Mickiewicza 30, PL-30059 Cracow, Poland}
\and
Z. Szvetelszky
\address{Faculty of Informatics, E{\"o}tv{\"o}s University,
H-1518 Budapest Pf. 120, Hungary}
\and
B. Szekf{\H u}
\address{BUTE FESS Department of Information and Knowledge Management,\\
P.O. Box: H-1521 Budapest Pf. 91, Hungary}
\and
K. Ku{\l}akowski
\footnote{\tt kulakowski@novell.ftj.agh.edu.pl}
\address{Faculty of Physics and Applied Computer Science,
AGH University of Science and Technology,
al. Mickiewicza 30, PL-30059 Cracow, Poland}
}
\maketitle


\begin{abstract}
We consider the average probability $X$ of being informed on a gossip in a given social
network. The network is modeled within the random graph theory of Erd{\H o}s and R\'enyi. 
In this theory, a network is characterized by two parameters: the size $N$ and the link 
probability $p$. Our experimental data suggest three
levels of social inclusion of friendship.
The critical value $p_c$, for which half of agents are informed, scales with the system size as $N^{-\gamma}$ with $\gamma\approx 0.68$.
Computer simulations show that the probability $X$ varies with $p$ as a sigmoidal curve.
Influence of the correlations between neighbors is also evaluated: with increasing clustering coefficient $C$, $X$ decreases.
 
\end{abstract}

\PACS{
87.23.Ge; 07.05.Tp
}


\section{Introduction}

Entering a new social group, we are vividly interested in all
kinds of non-formal contacts. They are necessary to interpret and qualify
properly all information we get: as relevant or marginal,
unique or commonly available, urgent or not so, etc. We are taught 
by evolutionary psychology \cite{buss} that this need
reflects the way of work of the human brain, as it has been formed 
during millions of years of evolution. This need forms then our today's relations
with people as well. As a consequence, it remains relevant for any 
social theory of human relations. That is why gossip appeared 
as an appealing catchword \cite{dunbar,db,emler,sabi,cs,lind}. Because of its 
roots noted above, theory of gossip can be seen as a part of evolutionary 
psychology. Once an evolutionary sociology emerges \cite{goog}, we will 
certainly find it there.

In sociophysics, we look at social sciences through a mathematical glass.
Being somewhat blind to hermeneutical analyzes, we look for determinism, 
structure and numbers. Such an attitude meets an old hope of sociologists
to deal with problems as well-defined and narrow as physicists have. (Invocations
to physics are quite frequent in old sociological textbooks \cite{homans,kempny}, 
to call only few examples.)  Sociology can meet with sociophysics in all cases where
the structure of society is of importance. By structure we mean a system
of mutual or directed connections between people. In a reductionistic approach,
such a system can be represented by a graph, where people are nodes and 
relations between people are links. Quite naturally, such a picture is a 
favorite tool in sociophysics.

Here we are going to use this mathematical representation to analyze the 
spreading of gossip. The starting point is the theory and experiment proposed
and performed recently by some of present authors \cite{zuza}. According to this 
theory, ``gossip is non-public information about knowable people and 
its primary attribute is proliferation. Gossiping is a communicative propensity 
characteristic of the human race manifesting itself in smaller communities''. 
Then, the person who is the subject of the gossip is known personally to the 
community. This fact makes the gossip interesting and this interest is the 
necessary condition of the gossip spreading. This ``semiprivate'' character makes
our case different from the theory of rumor by Galam \cite{galam}.
The experiment \cite{zuzabalasz} dealt with an interest in gossip about a 
known or knowable person in a web-based social network. As a result, three levels
of social inclusion have been found, which practically limited the gossip spreading.
As a sample of the questionnaire, answers were gathered to the following:
\begin{enumerate}
\item Would you tell about your girlfriend's new job to your friend? 
\item Would you tell about your girlfriend's new job to your friend's girlfriend? 
\item Would you tell about your girlfriend's new job to your friend's girlfriend's colleague?
\end{enumerate}
The percentage $q_i$ of positive answers varied from $q_1=100\%$ through $q_2=74.8\%$ till
$q_3=22.1\%$, respectively for 
questions 1, 2 and 3. After the third degree the results had shown 
a sharp decline \cite{zuzabalasz}.

These considerations led us to our main question, under what conditions a given gossip will
be known in the whole community? The above numbers $q_i$ ($i=1,2,3$) served us 
as probabilities, that the gossip will be told to people of 1-st, 2-nd and 3-rd level
of inclusion, defined by the questionnaire. At this point we are faced with the 
as-yet-unsolved problem, what is the structure of the social network? We have to admit that
the answer varies from one kind of network to another, one or another kind of social 
ties. In the literature of the subject, one can find arguments about different parameters
of social networks: size from a hundred to three hundreds and more \cite{dunbar,simmel,kill}, 
clustering coefficient \cite{ravasz}, strength of ties \cite{gran} and structure \cite{watts,pujol}. 
The results can depend 
also on whether complete networks or personal networks are investigated \cite{jackson}.
When we speak on friendly personal networks, the size of a typical group can decrease 
by at least one order of magnitude \cite{bunt}. Not entering these discussions,
here we attack the problem of gossip spreading in a model way, where the average number of
friends is a model parameter. Also, for simplicity we choose the random graph of 
Erd{\H o}s and R\'enyi \cite{erre} as a model of a social network. This selection should 
serve as a useful point of reference.

The goal of this paper is to calculate the probability that the gossip is known, averaged 
over the community members. Basically, the result is close to zero or one, except some range of 
the average number of friendship ties. This range can be seen as the range of a transition
between two phases: ``they do not know'' and ``they know''. However, even if the width of this
range eventually shrinks to zero in the limit of large networks, this limit is not relevant
for social sciences, where the quality of useful approximations does not necessarily increase with 
the system size.  

In next section we describe the model calculations and the results. Last section is devoted 
to their discussion.

\section{Calculations and results}

From noted above, the following model assumptions emerge:
\begin{enumerate}
\item  The set of nodes are those who ``know about'', and that is why they are willing to hear.
\item  The links join two nodes if they are friends.
\item  The linkage is random, as in the Erd{\H o}s--R\'enyi model.
\item  The question is to evaluate the size of the group who will know the information. 
\end{enumerate}

The detail is if the victim of the gossip is also a member of the network in which gossip is spread. 
The argument for
this assumption is in the questionnaire ``about your girlfriend's new job''. In this case 
the talker role is to be limited to the set of boyfriends of the girl.  
However, we assume that the girl can have more boyfriends, and then the number of talkers 
can be greater than one. 

In this case we have two parameters: $N$ (the number of nodes) and $p$ (the probability of a 
link of
friendship between randomly selected nodes). 
As we know from the theory of random networks \cite{doro}, the mean degree is $z=p(N-1)$.
The numbers $q_1$, $q_2$ and $q_3$ can be interpreted as weights in the average level of
being informed about a given gossip. All friends of the girl who got a job will know it with
probability one ($q_1=1$). This is a contribution $z$.
Their friends (each has $z-1$ still not informed) 
will know the gossip with probability $q_2$. This is a contribution $q_2z(z-1)$. Finally, 
consider friends of the friends (supposed they are not informed yet). If each friend has $z-1$ 
uninformed friends, the information will pass to them from the teller with, say, probability 
$q_3(z-1)z(z-1)$. Then, total level $X$ of being informed on the gossip would be 
\begin{equation}
\label{eq-X1}
X=\frac{q_1z+q_2z(z-1)+q_3z(z-1)^2}{N-1}.
\end{equation}
This is a function of $N$ and $p=z/(N-1)$.

This expression has some deficiency: 
in the random networks the probability that two ``friends'' of a node are also ``friends'' 
is $z/(N-1)=p$.
In the above calculation, we disregarded this possibility. 
Now we are going to include it. 
In the first zone, $z$ friends are informed with probability $q_1$.
Each has $(z-1)$ neighbors, $p$ of them are already informed.
Then, newly informed are only $(1-p)$ next neighbors, and their contribution will be 
$q_2z(z-1)(1-p)$.
How many still non-informed neighbors have these $z(z-1)(1-p)$ people?
The answer is that each has $(z-1)(1-p)$.
They will be informed by a teller with probability $q_3$.
Then, their contribution is $q_3z(z-1)^2(1-p)^2$.
The total formula is:
\begin{equation}
\label{eq-X2}
X=\frac{q_1z+q_2z(z-1)(1-p)+q_3z(z-1)^2(1- p)^2}{N-1}. 
\end{equation}
We note that still there are some assumptions left about the lack of correlations of further 
order, the arguments are somewhat heuristic, and valid only for small $p$.
However, $X$ obviously increases with $p$.
In the range where the formula is not valid (large $p$) we rely on a computer simulation.

For given victim of the gossip $i$ (one of $N$ nodes constituting 
network) and all $t(i)$ of its nearest neighbors (talkers) we evaluate 
the number $n_1(i)$ [$n_2(i)$] of paths of length $1$ [$2$] from all 
other nodes $j$ to talkers.
The probability that $j$-th node {\em is not} informed is
\begin{equation}
1-P_j=(1-q_2)^{n_1(i)} (1-q_3)^{n_2(i)}.
\end{equation}
Then, the level $X_i$ of being informed on the gossip for given victim 
$i$ is
\begin{equation}
X_i=\dfrac{t(i)+\sum_j P_j}{N-1},
\end{equation}
where summation goes over all $j\ne i$ and $j$ is not a talker.
The total level $X$ of being informed of the gossip is averaged over all 
possible victims of the gossip in the given network
\begin{equation}
X=\sum_{i=1}^N X_i/N.
\end{equation}
We carry out our simulation for the set of probabilities $q_1=1$, $q_2=0.748$ and $q_3=0.221$.

\begin{figure} 
\begin{center}
(a)
\includegraphics[scale=0.7]{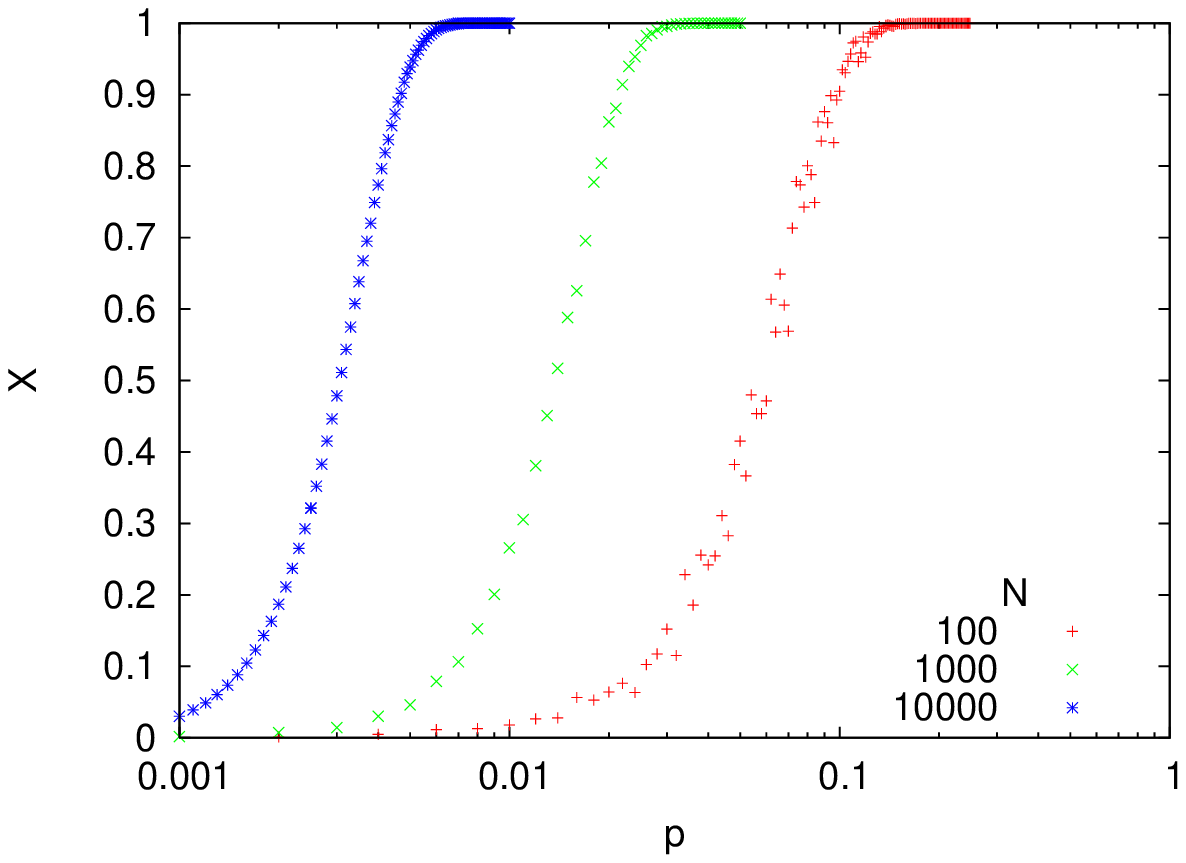}\\
(b)
\includegraphics[scale=0.7]{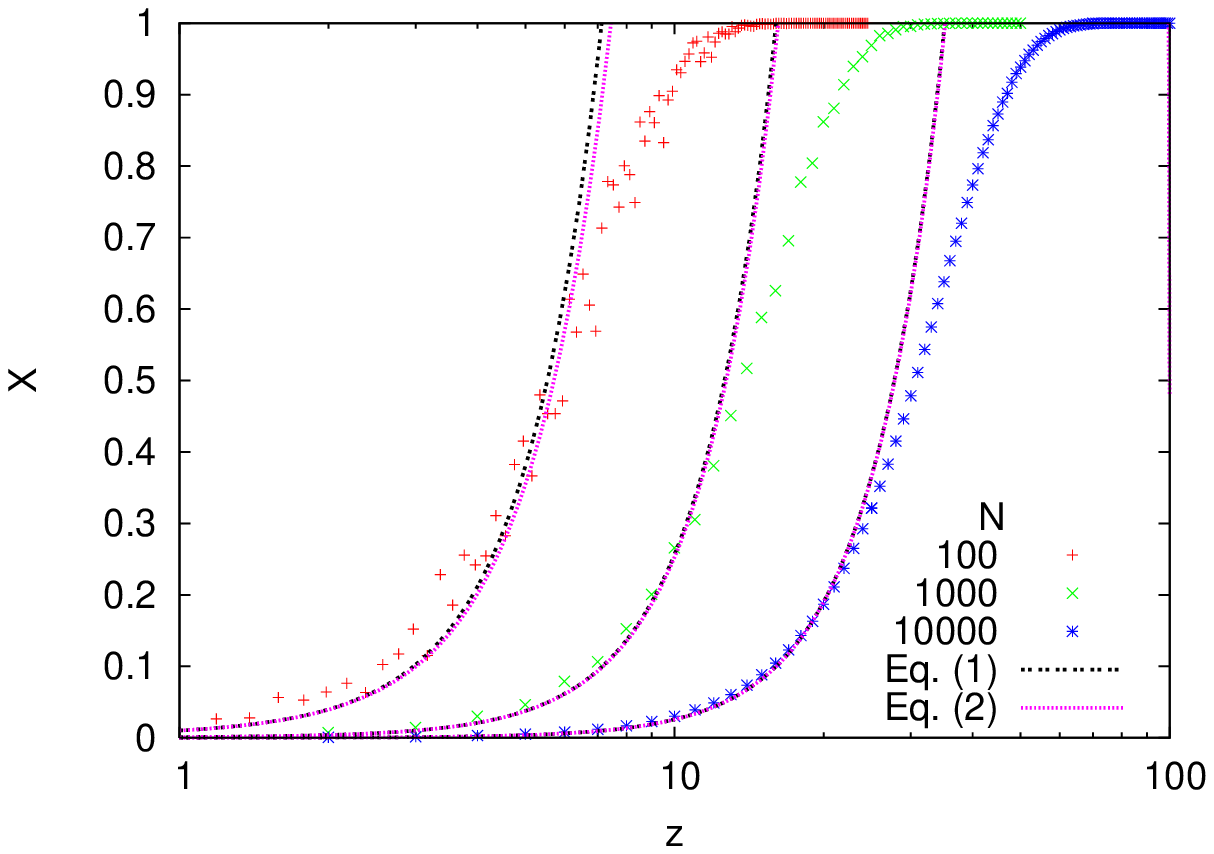}\\
\end{center}
\caption{\label{fig-Xp} Average probability $X$ of being informed on a gossip against (a) the probability $p$ and (b) mean node degree $z$.
Theoretical curves (Eqs. \eqref{eq-X1} and \eqref{eq-X2}) differ only slightly for $N=100$, but coincide for larger networks.
Their accordance with the simulation results improves for larger $N$, where the correlations between informed neighbors can be neglected with better accuracy.
}  
\end{figure}

It appears (Fig. \ref{fig-Xp}) that at some value of $p$, almost everybody will know the gossip.
This value of $p$ is however not strictly defined and it depends on the system size $N$.
For small $p$, both expressions (Eqs. \eqref{eq-X1} and \eqref{eq-X2}), for correlated and uncorrelated (i.e. random) case, work 
almost equally well.

\begin{figure}
\begin{center}
\includegraphics[scale=0.7]{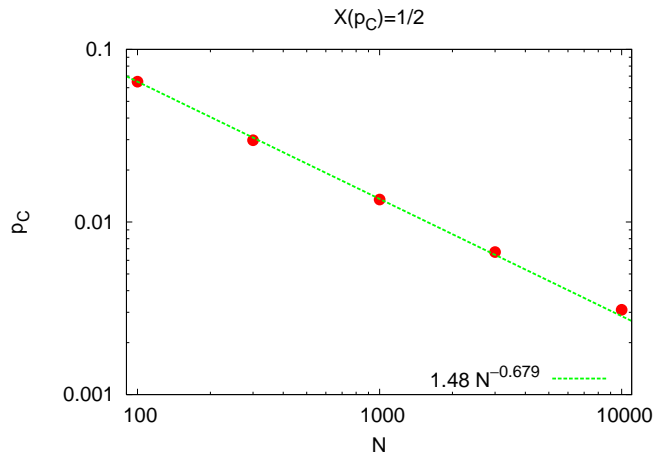}
\end{center}
\caption{\label{fig-PcN} Dependence of critical probability $p_c$ on the system size $N$.
The solid line shows the least square fit $p_c \propto N^{-\gamma}$ with $\gamma\approx 0.68$.}
\end{figure}

Motivated by tradition of statistical mechanics, we made an attempt to evaluate the probability 
$p_c$, where $X=1/2$. This $p_c$ can be seen as a critical value between the two phases remarked 
above, where ``they do not know'' for $p<p_c$ and ``they know'' for $p>p_c$. The size dependence of
$p_c$, i.e. $p_c(N)$, is shown in Fig. \ref{fig-PcN}. The results nicely fit a power law 
$p_c\propto N^{-\gamma}$. The exponent $\gamma$ slightly varies with the measured probabilities ($q_1$, $q_2$ and $q_3$);
it is 0.68 for the values of the probabilities (1.0, 0.748 and 0.221) used here, but 0.63 for (1.0, 0.7 and 0.25),
0.65 for (1.0, 0.75 and 0.25) and 0.66 for (1.0, 0.8 and 0.2).

We made also an attempt to evaluate the influence of the clustering coefficient $C$ on our results.
The coefficient $C$ is defined as the ratio of number of links between $k_i$ nearest neighbors of $i$-th site, divided by the maximal value of this number $(k_i(k_i-1)/2)$ and averaged over all sites of the network with more than one neighbor.
Our motivation comes from the suggestion \cite{wastr} that in social systems, the correlation 
is larger than for random case. The simulation is performed for $N=1000$ and $p=0.0135$, which is equal 
to $p_c(N=1000)$ for the random (i.e. uncorrelated) network. The clustering coefficient is increased by a rewiring procedure: a node is selected with at least $K_{\text{cut}}=3$ neighbors, and the link to one of its neighbor is cut; instead, it is added between two remaining neighbors. 

\begin{figure}
\begin{center}
\includegraphics[scale=0.7]{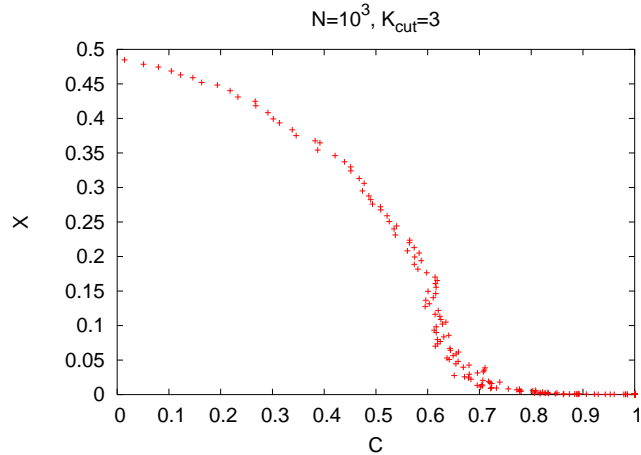}
\end{center}
\caption{\label{fig-XCC} Influence of the clustering coefficient $C$ on the level of gossip propagation $X$ for $N=1000$.
Increasing $C$ and keeping $p$ constant, we eventually get the network split in parts.}
\end{figure}

The result is that as $C$ increases, the average size $X$ of informed group decreases.
Example of this result is shown in Fig. \ref{fig-XCC}.
It is clear that in the case of larger $C$, information is transmitted more frequently 
within a small group.
On the contrary, its spread over the whole community is less effective.
This effect is parallel to the discussion in sociological literature, where links joining 
different 
compact groups (the so-called weak ties) are considered to be crucial for the information 
spreading \cite{gran}.

\section{Discussion}

When a social group is formed from the beginning, almost nobody knows anything about others.
Soon mutual ties are built and strengthen, and information starts to flow. In our picture,
this process can be interpreted as an increasing of the probability $p$ in time. The results
presented in Fig. 1 indicate, that the information carried by gossips increases initially with 
$p$ as a low degree polynomial. Gradually, the whole group becomes informed.  

Keeping the experimental values of $q_i$ constant, as we do, we can expect some characteristic distance $b$ 
from the victim to a member who is informed with probability, say, 1/2. (This distance is a graph characteristics
and it should not me mixed with the social distance, discussed elsewhere \cite{pab,my}). Certainly, this distance depends
on the numbers $q_i$, $i=1,2,3$. On the other hand, the diameter of the random network can be evaluated 
\cite{nwm} as $d=\ln N/\ln z$. At $p=p_c$ we can expect that $d=b$. Approximating $z$ by $Np$, we get
$b=\ln(N)/\ln(Np_c)$, i.e.
\begin{equation}
p_c=N^{1/b-1}
\end{equation}
Comparing this with our numerical result $p_c=N^{-\gamma}=N^{-0.68}$, we get $b$ close to $3.0$. Having in mind our
values of the probabilities $q_i$, we are not surprized with this distance. Reasonably enough, it agrees with
the interpretation of the experiment, given in our previous work entitled 
{\it Three levels of inclusion} \cite{zuzabalasz}. We conclude that the 
exponent $\gamma$ is not universal, but it depends on the probabilities $q_i$. 
With increasing $p$, the whole group is gradually dragged into the shell of radius $b$ around the victim.
Actually, the gossiping can be a good reason to enhance group ties.

In statistical mechanics, our results may be relevant for the percolation problem in random 
networks. It is known that large connected clusters appear for $p> 1/N$ \cite{erre,doro}.
Important difference is that in our case of gossip, we have one source of information.
In this aspect, the gossip spread can be compared to a spread of infection, e.g. in networks of sexual
interactions \cite{gon}.
Once we allow for a distribution of sources within the network, the problem of gossip
becomes alike to the family of problems, as bootstrap percolation \cite{adler} or diffusion 
percolation \cite{adah,asa}. It seems natural that these problems will find social applications,
similar to those \cite{gust} of standard percolation theory \cite{staah}. The bottleneck
here is the sociological experiment, which is much more difficult, costful and debatable
than computer simulations.

\section*{acknowledgments}
Three of the authors (Z.S, B.S. and K.K.) are grateful to the Organizers of the Eighth Granada Lectures in 2005, where this work was initialized, for their kind hospitality. 
Calculations were carried out in ACK\---CY\-FRO\-NET\---AGH.
The machine time on HP Integrity Superdome is financed by the Polish Ministry of Science and Education under grant No. MNiI/HP\_I\_SD/AGH/047/2004.


\begin{thebibliography}{99}
\bibitem{goog} Although the number of a search results in {\tt Google} can hardly serve as a measure of reality \cite{bagrow}, we note that it shows $86\times 10^3$ web pages with ``evolutionary psychology'' and only 1220 web pages with ``evolutionary sociology''.
For a comparison, ``physics'' reveals $334\times 10^6$ web pages.
These numbers suggest differences of scale.
\end{thebibliography}
\end{document}